\documentclass{article}

\usepackage{microtype}
\usepackage{graphicx}
\usepackage{subcaption}
\usepackage{booktabs}
\usepackage{url}
\usepackage{float}

\usepackage{hyperref}

\usepackage{acro}
\usepackage{tabularx}
\usepackage{multirow}

\graphicspath{{images/}}

\usepackage[preprint]{icml2026}

\newcommand*{\ApplyModerateFloatSpacing}{%
	\setlength{\floatsep}{9pt plus 2pt minus 2pt}%
	\setlength{\textfloatsep}{14pt plus 2pt minus 4pt}%
	\setlength{\intextsep}{9pt plus 2pt minus 2pt}%
	\captionsetup{skip=0.06in}%
}

\ApplyModerateFloatSpacing
\AtBeginDocument{\ApplyModerateFloatSpacing}

\usepackage{amsmath}
\usepackage{amssymb}
\usepackage{mathtools}
\usepackage{amsthm}

\newcommand*{\ApplyCompactDisplayMathSpacing}{%
	\setlength{\abovedisplayskip}{2pt plus 1pt minus 1pt}%
	\setlength{\belowdisplayskip}{2pt plus 1pt minus 1pt}%
	\setlength{\abovedisplayshortskip}{0pt plus 1pt}%
	\setlength{\belowdisplayshortskip}{1.5pt plus 1pt minus 1pt}%
	\setlength{\jot}{0.5pt}%
}

\ApplyCompactDisplayMathSpacing
\AtBeginDocument{\ApplyCompactDisplayMathSpacing}

\usepackage{algorithm}
\usepackage{algorithmic}

\usepackage{enumitem}
\setlist[itemize,enumerate]{%
	topsep=3pt plus 1pt minus 1pt,
	itemsep=1pt plus 0.5pt minus 0.5pt,
	parsep=0pt,
	partopsep=0pt,
}

\usepackage[capitalize,noabbrev]{cleveref}

\theoremstyle{plain}

\theoremstyle{definition}

\theoremstyle{remark}

\usepackage[disable,textsize=tiny]{todonotes}

\icmltitlerunning{Inductive Latent Context Persistence}

\DeclareAcronym{ai}{short = AI, long = artificial intelligence} 
\DeclareAcronym{ml}{short = ML, long = machine learning} 
\DeclareAcronym{ran}{short = RAN, long = radio access network} 
\DeclareAcronym{ue}{short = UE, long = user equipment} 
\DeclareAcronym{gnb}{short = gNB, long = next-generation Node\,B} 
\DeclareAcronym{enb}{short = eNB, long = evolved Node\,B} 
\DeclareAcronym{ho}{short = HO, long = handover} 
\DeclareAcronym{hof}{short = HOF, long = handover failure} 
\DeclareAcronym{cho}{short = CHO, long = conditional handover} 
\DeclareAcronym{daps}{short = DAPS, long = dual active protocol stack} 
\DeclareAcronym{ltm}{short = LTM, long = L2-triggered mobility} 
\DeclareAcronym{rrc}{short = RRC, long = radio resource control} 
\DeclareAcronym{rrm}{short = RRM, long = radio resource management} 
\DeclareAcronym{nr}{short = NR, long = New Radio} 
\DeclareAcronym{lte}{short = LTE, long = Long-Term Evolution} 
\DeclareAcronym{rsrp}{short = RSRP, long = reference-signal received power} 
\DeclareAcronym{rsrq}{short = RSRQ, long = reference-signal received quality} 
\DeclareAcronym{sinr}{short = SINR, long = signal-to-interference-plus-noise ratio} 
\DeclareAcronym{ssb}{short = SSB, long = synchronization signal block} 
\DeclareAcronym{ttt}{short = TTT, long = time-to-trigger} 
\DeclareAcronym{nlos}{short = NLOS, long = non-line-of-sight} 
\DeclareAcronym{los}{short = LoS, long = line-of-sight} 
\DeclareAcronym{xn}{short = Xn, long = Xn interface} 
\DeclareAcronym{oran}{short = O-RAN, long = open radio access network} 
\DeclareAcronym{ric}{short = RIC, long = RAN intelligent controller} 
\DeclareAcronym{xapp}{short = xApp, long = extended application} 
\DeclareAcronym{gnn}{short = GNN, long = graph neural network} 
\DeclareAcronym{gat}{short = GAT, long = graph attention network} 
\DeclareAcronym{hgt}{short = HGT, long = heterogeneous graph transformer} 
\DeclareAcronym{gru}{short = GRU, long = gated recurrent unit} 
\DeclareAcronym{lstm}{short = LSTM, long = long short-term memory} 
\DeclareAcronym{vae}{short = VAE, long = variational autoencoder} 
\DeclareAcronym{vgae}{short = VGAE, long = variational graph autoencoder} 
\DeclareAcronym{ema}{short = EMA, long = exponential moving average} 
\DeclareAcronym{kl}{short = KL, long = Kullback--Leibler} 
\DeclareAcronym{mse}{short = MSE, long = mean-squared error} 
\DeclareAcronym{ci}{short = CI, long = confidence interval} 
\DeclareAcronym{zk}{short = ZK, long = Zero-Knowledge} 
\DeclareAcronym{ilcp}{short = ILCP, long = Inductive Latent Context Persistence} 
\DeclareAcronym{drx}{short = DRX, long = discontinuous reception} 
\DeclareAcronym{tdd}{short = TDD, long = time-division duplex} 
\DeclareAcronym{kpi}{short = KPI, long = key performance indicator} 
\DeclareAcronym{nextg}{short = NextG, long = Next-Generation Wireless} 
\DeclareAcronym{mlp}{short = MLP, long  = multilayer perceptron}
\DeclareAcronym{rlf}{short = RLF, long  = radio link failure}

\begin{document}
	
	\twocolumn[
	\icmltitle{Inductive Latent Context Persistence:
		Closing the Post-Handover Cold Start in 6G Radio Access Networks}
	
	\icmlsetsymbol{equal}{*}
	
	\begin{icmlauthorlist}
		\icmlauthor{Anubhab Banerjee}{equal,nokia}
		\icmlauthor{Daniyal Amir Awan}{equal,nokia}
	\end{icmlauthorlist}
	
	\icmlaffiliation{nokia}{Nokia Solutions and Networks GmbH \& Co.\ KG, Munich, Germany}
	
	\icmlcorrespondingauthor{Anubhab Banerjee}{anubhab.1.banerjee@nokia.com}
	\icmlcorrespondingauthor{Daniyal Amir Awan}{daniyal.awan@nokia.com}
	
	\icmlkeywords{AI-native RAN, 5G-Advanced, 6G, mobility prediction, handover, graph neural networks, in-context transfer, O-RAN xApp}
	
	\vskip 0.3in
	]
	
	\printAffiliationsAndNotice{\icmlEqualContribution}
	
	\begin{abstract}
In modern \acp{ran} rule-based \ac{ho} decisions (e.g., A3/A5) depend on \ac{ue} measurements only, resulting in different \ac{ho} decisons for \acp{ue} who are in the same location.
To overcome it, \acp{gnn} based methods have been proposed to improve \ac{ho} \acp{kpi} using more information than just the measurements. 
However, existing recurrent or graph-based methods discard the per-\ac{ue} recurrent state at \ac{ho} and reinitialize it at the target \ac{gnb} -- losing useful measurement and mobility history as part of the overall context, forcing the target-side model to rebuild the state from post-\ac{ho} measurements only.
We address this post-\ac{ho} cold start problem with \ac{ilcp}, a learned latent synchronization mechanism that compresses the source-side per-\ac{ue} recurrent state, transports it over the standard 3GPP \ac{xn} as a 128-byte payload, and adapts it to the target \ac{gnb} state space at \ac{ho}. We model the \ac{ran} as a dynamic heterogeneous graph, allowing the model to treat \ac{ue} nodes, \ac{gnb} nodes, measurement edges, and \ac{xn} neighbor edges separately.
On the Vienna 4G/5G drive-test, \ac{ilcp} eliminates ping-pong \acp{ho} in the test split with 0.0\% compared with 6.5\% for the otherwise identical but no-transfer baseline and 22.6\% for a Transformer baseline. It also achieves a $+5.1$\,pp average post-\ac{ho} accuracy gain, with a peak gain of $+13.3$\,pp, over the no-transfer baseline in the 50--250\,ms post-\ac{ho} window. On a single NVIDIA GTX~1080 (8\,GB), ILCP runs end-to-end at $7.7$\,ms p99 per handover decision.
Under measurement perturbations including shadow fading, \ac{nlos} blockage, and \ac{ssb}-burst sparsity, robustly trained \ac{ilcp} keeps \ac{hof} in the 10--13\% and exposes sub optimality of relying only on measurements. In the same fixed-reference-label setting, the 3GPP A3/A5 rule which relies only on measurements increases from 1.1\% \ac{hof} on the unperturbed trace to 57--65\% under perturbed measurements. 
\end{abstract}

	\section{Introduction}
\label{sec:intro}
Modern dense \acp{ran} deployments may cover a relatively small area with many overlapping capacity and coverage cells, and a \ac{ue} moving through these should be handed over from one cell to another to maintain a strong radio connection with the network. Obviously, the quality of these \acp{ho} is an important \ac{kpi} of the network deployment and \ac{ho} control. In dense networks, with fast moving \acp{ue}, mobility performance is critical. A missed \ac{ho} may cause a \ac{rlf} and possibly a multi-second outage, while an error-prone \ac{ho} between two adjacent cells (e.g., a \emph{ping-pong} \ac{ho}) wastes control-plane signaling in addition to data loss \citep{tgpp_38_300}. Due to these reasons, work in 3GPP has produced a family of event-triggered rules (A3/A5, \ac{cho}, \ac{daps}, \ac{ltm}) that generally decide \acp{ho} from two simple parameters: a hysteresis margin, and a \ac{ttt}, applied to filtered or averaged \ac{rsrp} measurements \citep{tgpp_38_331}. These rules are simple, robust and may be tuned per deployment. However, they do not do predictive extrapolations and they make decisions based on current measurement evidence only which are often imperfect due to a variety of reasons. 

With increasing usage of DNNs and large computing, past work studied recurrent and graph-attention models trained on real drive-test traces and network topologies which match or beat the rule-based algorithm on next-cell accuracy at the \ac{ho} instant, while also extrapolating into the future where rule-based events fail \citep{ozturk2019novel, wickramasuriya2017base, lin2019data, hu2020heterogeneous, velickovic2018graph}. The learning process consumes a per-step network information stream including topologies and measurements, a recurrent or self-attentive temporal module integrating the stream into a hidden or embedded state $h_u(t)$ for each \ac{ue} $u$, and a downstream head using that hidden state to produce a probability-type estimate over next possible cell(s). In short, a \ac{gnn} captures the spatial geometry, the recurrent network module captures the temporal dynamics, and the head network ranks the next-cell candidates.


The hidden state $h_u(t)$ summarizes a \ac{ue}'s local mobility-relevant context comprising of signal measurements from the serving \ac{gnb}. Now, let us consider two critical issues: (i) a \ac{ho} transfers the \ac{ue} to a different \ac{gnb} that has its own model state and local context. Then, the embedding $h_u(t)$, although representative of the context in the source \ac{gnb}, may be seen as no longer too relevant for the target \ac{gnb} after the \ac{ho}. (ii) currently, the standardized Xn {\sc HANDOVER\,REQUEST} carries \ac{rrc} state, security context, and \ac{ue} capabilities \citep{tgpp_38_423} as part of knowledge-transfer before \acp{ho}. Therefore, the target \ac{gnb} will start $h_u(t)$ from scratch and must rebuild a useful $h_u$ from the few measurements it has since the \ac{ho}. We call this cold restart of the embedding as \emph{Zero-Shot Cold Start} that wastes the historical learned representation of UE context and its usefulness in predicting the next best cell and avoiding an \ac{rlf} or ping-pong. 

To tackle these two issues above and drawbacks of the state-of-the-art, we treat the per-\ac{ue} recurrent state as a portable network context which can be transferred (as part of overall knowledge-transfer for the \ac{ho}) between \acp{gnb}. To address the practical issue of Xn message limits, we show that a 128-byte differential update is sufficient to preserve the predictive quality of a 128-dimensional \ac{gru} state across the \ac{ho} boundary. To achieve this, our proposed \ac{ilcp} protocol compresses the hidden state with a $\beta$-\ac{vae}, transports it on the standard 3GPP Xn interface, and projects it onto the target \ac{gnb}'s state space at the moment of handover via a learned, gated \ac{mlp}. 

We validate \ac{ilcp} on a real drive-test traces from Vienna 4G/5G against five baselines using the original rule-based based next-cell decisions as the ground-truth ``best cell". Against the 3GPP A3/A5 rule, we do a comparison under measurement impairments expected of a real network to argue that relying on measurements and fixed rules alone results in sub optimal \ac{ho} decisions.

\textbf{Contributions}: \textbf{(i)} We formalize the Zero-Shot Handover Cold Start as inductive domain shift on a dynamic heterogeneous graph in which the target \ac{gnb} must infer a temporal hidden state without access to the measurement history (Section~\ref{sec:problem}). \textbf{(ii)} We introduce \ac{ilcp} (Section~\ref{sec:method}), a differential latent synchronization protocol whose 128-byte payload piggy-backs on the existing 3GPP \ac{xn} handover messages. \textbf{(iii)} On the Vienna trace \ac{ilcp} improves post-handover next-cell accuracy by $+5.1$\,pp average (peak $+13.3$\,pp), eliminates ping-pongs entirely ($0.0\%$ vs.\ $6.5\%$ for the Zero-Knowledge transfer baseline and $22.6\%$ for a Transformer baseline). \textbf{(iv)} Under realistic measurement impairments robustly trained \ac{ilcp} holds \ac{hof} at $10\!\!-\!\!13\%$ while the 3GPP A3/A5 rule collapses from $1.1\%$ to $57\!\!-\!\!65\%$ showing the rule-based limitation of idealized noise-free traces.

	\section{Related Work}
\label{sec:relatedwork}

\textbf{Predictive mobility management}: Classical \ac{ml} approaches to handover prediction range from random forests on \ac{rsrp} features to recurrent networks consuming time-step measurements \citep{ozturk2019novel, wickramasuriya2017base, lin2019data, banerjee2022trust, banerjee2021toward, banerjee2020game}. 
More recent work applies \acp{gnn} to model the \ac{ue}--cell topology explicitly \citep{velickovic2018graph, hu2020heterogeneous, hamilton2017inductive}, and \ac{cho} / \ac{daps} variants \citep{tgpp_38_300, tgpp_38_331} add procedural support for early target preparation. 
None of these works persists learned representations across the handover boundary; the predictor at the target \ac{gnb} starts from a re-initialized state.

\textbf{In-context transfer of learned state}: Federated and split-learning systems \citep{mcmahan2017communication, kairouz2021advances} share \emph{parameter} updates between training rounds; our work differs in that the unit of transfer is a single \ac{ue}'s recurrent state at the moment of handover, the transport latency budget is on the order of the handover preparation window (${\approx}100$\,ms), and the target-side projection is a learned operator. \ac{ilcp} is therefore closer to \emph{in-context} transfer of episodic memory than to federated optimization, and is complementary to continual-learning methods \citep{kirkpatrick2017overcoming} that mitigate forgetting in a single per-\ac{gnb} model. 
The compressor is a $\beta$-\ac{vae} \citep{higgins2017beta, kipf2016variational} optimized end-to-end with the downstream candidate-set scoring loss, keeping with the information-bottleneck view of representation learning \citep{tishby2015deep}. The realignment of source-side and target-side state spaces is reminiscent of CORAL and Subspace Alignment \citep{sun2016deep, fernando2013unsupervised}, but specialized to the heterogeneous-graph plus recurrent-state setting.

\textbf{Robust learning for telecom data}: For comparison with A3/A5 under measurement inaccuracies, our mixed-perturbation training recipe is closely related to AugMix and consistency training \citep{hendrycks2020augmix, xie2020unsupervised, miyato2018virtual}.

	\section{System and Problem Formulation}
\label{sec:problem}

In this section we formulate the overall system and learning problem. See Figure~\ref{fig:system_diagram} for an end-to-end system diagram. 
For readability purposes and because of page constraint, Figure~\ref{fig:system_diagram} has been put in Appendix~\ref{sec:appx-system-diagram}.

\textbf{Dynamic heterogeneous \acs{ran} graph}: At any time instant $t$, we represent the radio access network as a dynamic heterogeneous graph
\begin{equation}
    G(t)=\bigl(V_{\mathrm{UE}} \cup V_{\mathrm{gNB}},\;
    E_{\mathrm{meas}}(t)\cup E_{\mathrm{Xn}}\bigr)
\end{equation}
Here, $V_{\mathrm{UE}}$ denotes the set of currently active \acp{ue}, and $V_{\mathrm{gNB}}$ denotes the set of deployed base stations. The edge set contains two relation types:
\begin{itemize}
    \item $E_{\mathrm{meas}}(t)\subseteq V_{\mathrm{UE}}\times V_{\mathrm{gNB}}$, the time-varying set of \ac{ue}--cell measurement edges, with per-edge attributes such as \ac{rsrp}, \ac{rsrq}, and \ac{sinr};
    \item $E_{\mathrm{Xn}}\subseteq V_{\mathrm{gNB}}\times V_{\mathrm{gNB}}$, the deployment-static set of neighboring \acp{gnb} connected via the \ac{xn} interface.
\end{itemize}
For each \ac{ue} $u\in V_{\mathrm{UE}}$, we denote its serving cell at time $t$ by $c_u(t)\in V_{\mathrm{gNB}}$.

\textbf{Prediction model}: Our model comprises following:
\begin{enumerate}
    \item Given the graph snapshot $G(t)$, a heterogeneous-attention encoder $f_\theta$ produces spatial embeddings for the \acp{ue} and candidate cells. In particular, for each \ac{ue} $u$ we compute
    \begin{equation}
        x_u(t)=f_\theta\!\left(G(t),u\right)\in\mathbb{R}^{d_x},
    \end{equation}
    and for each candidate cell $c_k$ we denote its embedding by $e_{c_k}\in\mathbb{R}^{d_x}$. The embedding $x_u(t)$ summarizes the \ac{ue}'s current local radio and graph context, including measured link quality to visible cells, the current serving-cell relation, neighboring candidate cells, and the connectivity structure induced by measurement and \ac{xn} edges.

    \item A recurrent module $r_\phi$ aggregates the per-step \ac{ue} embeddings over time to form a temporal state
    \begin{equation}
        h_u(t)=r_\phi\!\bigl(h_u(t-1),x_u(t)\bigr)\in\mathbb{R}^{d_h},
    \end{equation}
    which encodes recent temporal information from the \ac{ue}'s measurement and mobility evolution, such as changes in link qualities, visible candidates, and serving-cell associations over the recent past.

    \item At time $t$, we consider only the $K$ visible candidate cells with the highest current \ac{rsrp} to be the serving cell at horizon $\Delta$. For each candidate cell $c_k$, a candidate-set scoring function $g^{(\Delta)}_\psi$ assigns the scalar score
    \begin{equation}
        s^{(\Delta)}_{u,k}(t)
        =
        g^{(\Delta)}_\psi\!\left(
        \left[
            h_u(t),\,
            e_{c_k},\,
            h_u(t)\odot e_{c_k}
        \right]
        \right),
        \label{eq:candidate-score}
    \end{equation}
    where $\odot$ denotes element-wise multiplication. The term $h_u(t)\odot e_{c_k}$ gives the scorer an explicit dimension-wise interaction between the \ac{ue} temporal state and the candidate-cell embedding. The predicted serving cell at horizon $\Delta$ is then
    \begin{equation}
        \hat c_u(t+\Delta)
        =
        \arg\max_{k\in\{1,\dots,K\}}
        s^{(\Delta)}_{u,k}(t).
        \label{eq:candidate-prediction}
    \end{equation}
\end{enumerate}

\textbf{Zero-shot handover cold start problem}: Consider a handover of \ac{ue} $u$ at time $t^\star$, where the serving cell changes from the source cell $c_u(t^\star\!-)$ to the target cell $c_u(t^\star\!+)$. Immediately before handover, the recurrent state $h_u(t^\star\!-)$ has accumulated information from the pre-handover observation sequence. Immediately after handover, the target side only has access to the post-handover measurement edges. If we do not transfer state across the handover, the recurrent module must restart from an initialization $h_u^{\mathrm{init}}$, and therefore loses the useful temporal context accumulated before handover. We quantify the resulting cold-start penalty by comparing a \emph{warm} predictor, which receives oracle access to the pre-handover hidden state, with a \emph{cold} predictor, which restarts from $h_u^{\mathrm{init}}$:
\begin{equation}
	\begin{aligned}
		\mathcal{L}_{\mathrm{cold}}
		&= \sum_{\substack{(t^\star,u)\\ \mathrm{handovers}}}
		\Bigl[
		\ell\!\left(\hat c_u^{\mathrm{cold}}(t^\star),\, c_u^\star(t^\star)\right)
		\\
		&\qquad - \ell\!\left(\hat c_u^{\mathrm{warm}}(t^\star),\, c_u^\star(t^\star)\right)
		\Bigr],
	\end{aligned}
	\label{eq:cold-start-cost}
\end{equation}
where $c_u^\star(t^\star)$ denotes the observed serving cell in the unperturbed reference trace, $\ell$ is the 0/1 prediction loss, $\hat c_u^{\mathrm{cold}}$ denotes prediction with re-initialized target-side state, and $\hat c_u^{\mathrm{warm}}$ denotes prediction with oracle access to the pre-handover hidden state. This gap captures the cost of the inductive domain shift introduced by handover.

\textbf{Objective and Method}: Our goal is to reduce the gap between the cold and warm predictors at the target \ac{gnb}, using only a constant-size message delivered over the existing \ac{xn} interface under realistic latency constraints, in the next section we present our proposed \ac{ilcp} method. \ac{ilcp} maps the pre-handover recurrent state $h_u(t^\star\!-)$ to a compact latent $z_u$, transports $z_u$ over \ac{xn}, and reconstructs an adapted target-side state $h_u^{\mathrm{new}}$ that can act as a substitute for the re-initialized state $h_u^{\mathrm{init}}$ after handover. 
We fix the candidate-set size to $K=8$ as a typical operational value. Note that the scoring function $g^{(\Delta)}_\psi$ is applied independently to each candidate and shared across the $K$ candidates. The method therefore does not depend on a fixed global parameterization of cell identifiers and can handle candidate-set changes caused by topology evolution.

\section{Inductive Latent Context Persistence}
\label{sec:method}
We implement \ac{ilcp} as a compact (128-Byte), learned synchronization protocol that can be carried over the existing 3GPP \ac{xn} interface. 
We train the full model for the downstream candidate-set prediction task. 
For each \ac{ue} $u$ at time $t$, we convert the candidate scores from Section~\ref{sec:problem} into a probability distribution over the visible candidate set:
\begin{equation}
p_\Theta^{(\Delta)}\!\bigl(c_k \mid u,t\bigr)
=
\frac{
    \exp\!\left(s^{(\Delta)}_{u,k}(t)\right)
}{
    \sum\limits_{j=1}^{K}
    \exp\!\left(s^{(\Delta)}_{u,j}(t)\right)
},
\label{eq:candidate-prob}
\end{equation}
where $\{c_1,\dots,c_K\}$ is the visible candidate set at time $t$, and $s^{(\Delta)}_{u,k}(t)$ is defined in Eq.~\eqref{eq:candidate-score}. We then optimize the downstream prediction objective
\begin{equation}
\mathcal{L}_{\mathrm{pred}}^{(\Delta)}
=
-\sum_{(u,t)}
\log p_\Theta^{(\Delta)}\!\bigl(c_u(t+\Delta)\mid u,t\bigr),
\label{eq:pred-loss}
\end{equation}
where $c_u(t+\Delta)$ is the ground-truth serving cell at horizon $\Delta$. To this end, we jointly train the heterogeneous-graph encoder $f_\theta$, the recurrent module $r_\phi$, the candidate-set scorer $g^{(\Delta)}_\psi$, the latent compressor $\mathcal{E}_\phi$, the latent decoder $\mathcal{D}_\psi$, and the target-side projection block comprising the gate network and the \ac{mlp} in Eq.~\eqref{eq:ilcp-update}. See Figure~\ref{fig:system_diagram} for an end-to-end system diagram. 

\textbf{Heterogeneous-graph backbone}: We encode each graph snapshot $G(t)$ with three \ac{hgt} layers \citep{hu2020heterogeneous} and use 4 attention heads per layer. 
We use type-specific attention parameters for \ac{ue}$\leftrightarrow$\ac{gnb} measurement edges and \ac{gnb}$\leftrightarrow$\ac{gnb} \ac{xn} edges, which preserves the asymmetry between instantaneous radio-measurement relations and deployment-static topology relations. By contrast, a relation-agnostic GAT \citep{velickovic2018graph} averages across both edge classes and loses about 30\,pp on our main accuracy metric (see GAT-Temporal in Table~\ref{tab:headline-vienna}).

\textbf{Per-\ac{ue} recurrent module}: We instantiate the recurrent module $r_\phi$ as a \ac{gru} \citep{cho2014learning} with hidden dimension $d_h=128$, and update the temporal state as
\begin{equation}
    h_u(t)=r_\phi\!\bigl(h_u(t-1),x_u(t)\bigr)=\mathrm{GRU}\bigl(h_u(t-1),x_u(t)\bigr).
\end{equation}
This state encodes recent temporal information from the \ac{ue}'s measurement and mobility evolution. In state-of-the-art, $h_u(t^\star\!-)$ is discarded after every handover whereas we preserve and transfer it through the steps below.

\textbf{$\beta$-\ac{vae} compressor}: We compress each recurrent state $h_u\in\mathbb{R}^{128}$ into a 32-dimensional latent vector
\[
z_u=\mathcal{E}_\phi(h_u)
\]
with an end-to-end-trained $\beta$-\ac{vae} \citep{higgins2017beta, kipf2016variational}, which yields a 128-byte FP32 payload. 
We train the compressor with
\begin{equation}
    \mathcal{L}_{\mathrm{VAE}}
    = \mathbb{E}_{q_\phi(z|h)}\!\bigl[\log p_\psi(h\mid z)\bigr]
    - \beta\,\mathrm{KL}\!\bigl(q_\phi(z\mid h)\,\|\,p(z)\bigr),
\end{equation}
with a suitable value of $\beta$, and optimize it jointly with Eq.~\eqref{eq:pred-loss}. Joint training makes the compressor task-aware such that it preserves those dimensions of $h_u$ that matter for the candidate scorer after an \ac{ho} has taken place recently, and compresses the rest. 

\textbf{Latent transport over the \ac{xn} interface}: At handover time $t^\star$, we append $z_u$ as an optional information element to the {\sc HANDOVER\,REQUEST} message defined in TS\,38.423 \citep{tgpp_38_423}. 
We note that the resulting 128-byte payload fits within existing message-size budgets.

\textbf{Gated projection at the target \ac{gnb}}: When the target \ac{gnb} receives the payload, we first decode it into
\[
\tilde h_u=\mathcal{D}_\psi(z_u).
\]
We then compute the target-side spatial embedding $x_u^{\text{new}}$ from the new local graph context and combine them through
\begin{equation}
    h_u^{\text{new}}
    =
    \mathrm{LN}\!\Bigl(
        \tilde h_u
        + \gamma\!\odot\!\mathrm{MLP}\!\bigl([\tilde h_u, x_u^{\text{new}}]\bigr)
    \Bigr),
    \label{eq:ilcp-update}
\end{equation}
where,
\[
\gamma=\sigma\!\bigl(g_\theta([\tilde h_u, x_u^{\text{new}}])\bigr)
\]
is a learned sigmoid gate and $\mathrm{LN}$ denotes LayerNorm. We treat $\tilde h_u$ as a transferred prior and use the gated \ac{mlp} term to adapt that prior to the new serving-cell context. The gate $\gamma$ controls how strongly we correct the transferred state using the new target-side observation, while LayerNorm keeps the combined representation well scaled and prevents either from dominating purely because of norm drift between source-side and target-side embedding spaces. During training, we apply inbound dropout to $\tilde h_u$ with rate 0.2. This keeps the projection block useful even when the transferred state is incomplete or partially corrupted.


\textbf{Supervised and Robust training}:
\label{para:method-robust}
Before moving to the experiments, we clarify the supervised training used in this paper. We use the serving-cell sequence observed in the trace as the reference label and assume, for the purpose of supervised training, that this sequence represents the desired next-cell target. This is an experimental assumption as the ``best cell" cannot be observed in the network trace. In fact, the logged serving cell is the decision realized by the deployed handover logic, which is typically A3/A5-like, and it need not always be the operationally best cell. This assumption also affects the comparison with A3/A5 on the unperturbed trace. If the label is itself aligned with A3/A5 behavior, then A3/A5 has an inherent advantage on exact-match metrics. For this reason, we interpret the unperturbed-trace comparison with care and use additional metrics, such as ping-pong rate and post-handover recovery, to evaluate whether a method produces stable mobility behavior. To model unavoidable variability in \ac{ue} measurements, and to expose limitations of A3/A5 rules in recovering an optimal best serving cell consistently, we also train robust versions of both \ac{ilcp} and the Zero-Knowledge baseline. At each step, we sample the input measurement stream from a mixture of the unperturbed reference trace, shadow-fading-corrupted \ac{rsrp} ($\sigma_s\sim\mathcal{U}\{6,12\}$\,dB), random \ac{nlos} blockage, and \ac{ssb}-burst sub-sampling. We keep the supervision target fixed to the serving cell observed on the unperturbed reference trace, so the model learns to recover the same reference decision under impaired measurements rather than imitate impairment-induced fluctuations. Finally, we remark that the experiment results are somewhat limited due to the limitations of the available trace data such as fixed topology (so that denser networks cannot be considered) and a limited number of actual \acp{ho}. In the future extension of this work, we will use network simulators to produce traces with optimal serving cells (in some sense), i.e., not relying on A3/A5 decisions, and demonstrate the full potential of our method in different network topologies. We will also use a multitude of node features that help predict the ``operationally best cell".

	\section{Experiments}
\label{sec:experiments}
We evaluate the setup of Section~\ref{sec:method} on the Vienna trace~\citep{vienna_dataset}, using the baselines and metrics below.
It is a multi-cell, multi-tier urban trace with dense cell overlap, 31 handover events in the held-out test split, and per-step measurements at 100 Hz. 
Unless noted, main-body results use the unperturbed trace; measurement-impairment studies follow Appendix~\ref{sec:exp-perturb} under the fixed-reference-label protocol described in Section~\ref{sec:method}.

\textbf{Baseline Methods}: We compare \ac{ilcp} against five baseline methods and one rule-based handover rule.
\begin{enumerate}
    \item \textit{Zero-Knowledge} \ac{hgt} is an otherwise identical \ac{hgt}+\ac{gru} prediction method but which re-initializes the per-\ac{ue} recurrent state at every \ac{ho}. This baseline isolates the effect of cross-handover state persistence because it shares the same graph backbone, recurrent module, and candidate-set scorer as \ac{ilcp}.

    \item \textit{GAT-Temporal} uses a single-relation \ac{gat}~\citep{velickovic2018graph} with the same hidden dimension, followed by the same \ac{gru}. This baseline tests the value of relation-specific heterogeneous attention.

    \item \textit{Transformer-Temporal} uses the same \ac{hgt} backbone followed by a Transformer encoder~\citep{vaswani2017attention} over a sliding window of length $T=16$. This baseline compares recurrent state accumulation with windowed self-attention.

    \item \textit{LSTM} is a direct sequence model over the raw per-step measurement vector~\citep{hochreiter1997long}, without the heterogeneous graph backbone.

    \item \textit{A3/A5} is the standard 3GPP event-triggered handover rule~\citep{tgpp_38_300,tgpp_38_331}, with hysteresis $H$ and \ac{ttt} $T$ set to default values.
\end{enumerate}

All neural-network-based methods use the architectural hyperparameters in Table~\ref{tab:hparams} of Appendix~\ref{sec:appx-setup}. Unless stated otherwise, we report the mean over 1000 bootstrap resamples of the test split, together with 95\% \acp{ci}; the resampling procedure is given in Appendix~\ref{sec:appx-bootstrap}.

\textbf{Metrics}: In the following, Acc@$t=0$ denotes next-cell accuracy when we evaluate the model at the \ac{ho} instant $t^\star$. The target is the future serving cell (selected using fixed network logic) at the prediction horizon $\Delta$, i.e., $c_u(t^\star+\Delta)$. After an \ac{ho}, Acc@$\delta$ denotes next-cell accuracy when we evaluate the model $\delta$ steps after the \ac{ho} instant. For an \ac{ho} at time $t^\star$, this means that the model is evaluated at $t^\star+\delta$ and its output is compared with the ground-truth future serving cell $c_u(t^\star+\delta+\Delta)$. Plotting Acc@$\delta$ as a function of $\delta$ shows how quickly the target-side model recovers useful context after \ac{ho}, and therefore makes the post-\ac{ho} cold-start gap visible. In the following, \ac{hof} is the fraction of \ac{ho} events for which the predicted target cell at the \ac{ho} instant is incorrect, while Ping-pong rate (PP) represents problematic \ac{ho} that are reversed within a short time-frame (for example, 500 ms).
\subsection{Overall Mobility-Prediction Performance}
\label{sec:exp-overall}

Table~\ref{tab:headline-vienna} summarizes the main mobility-prediction metrics on Vienna 4G/5G with bootstrap 95\% \acp{ci}. We observe that \ac{ilcp} eliminates ping-pongs entirely with PP $=0.0\%$, compared with $6.5\%$ for the otherwise identical Zero-Knowledge baseline and $22.6\%$ for the Transformer baseline. At the handover instant, \ac{ilcp} reaches $83.9\%$ Acc@$t=0$, while the Zero-Knowledge baseline reaches $87.1\%$. This difference is within the reported confidence intervals and, as shown in Section~\ref{sec:exp-postho}, the ordering changes in the early post-handover window. The sequence-only baselines perform substantially worse. LSTM reaches only $12.9\%$ Acc@$t=0$, and GAT-Temporal reaches $22.6\%$. This indicates that the heterogeneous graph structure carries important predictive information. In particular, relation-specific modeling of \ac{ue}-to-cell measurement edges and \ac{xn} neighbor edges is more useful than recurrence alone. 
The A3/A5 rule reaches $100\%$ on the clean trace trivially because the \ac{ho} labels are tied to the rule-based \ac{ho} behavior so the comparison here is misleading and only shown for sanity check. The \ac{ilcp} does not match the A3/A5 rule even if it is trained to predict the observed serving cells because, in contrast to A3/A5, \ac{ilcp} uses \ac{ue}'s network graph context summarized by its embedding which also includes measurements but does not solely rely on them. We show this advantage in Appendix~\ref{sec:exp-perturb} where we show the limitation of A3/A5 and other baselines under practical measurement variations and impairments. 

\begin{table}[t]
    \centering
    \caption{Headline mobility-prediction metrics on the Vienna 4G/5G drive-test (test split, 31 handovers)}
    \label{tab:headline-vienna}
    \small
    \setlength{\tabcolsep}{3pt}
    \begin{tabular}{lcccc}
        \toprule
        Method & Acc@t=0 & HOF & PP & Ovr \\
        \midrule
        \textbf{ILCP (ours)} & 83.9\,\scriptsize{(71--94)} & 16.1\,\scriptsize{(3--32)} & 0.0\,\scriptsize{(0--0)} & 74.1 \\
        \acs{zk}-HGT & 87.1\,\scriptsize{(74--97)} & 12.9\,\scriptsize{(3--26)} & 6.5\,\scriptsize{(0--16)} & 75.6 \\
        GAT-Temporal & 22.6\,\scriptsize{(10--39)} & 77.4\,\scriptsize{(61--90)} & 61.3\,\scriptsize{(45--77)} & 19.3 \\
        Transformer & 77.4\,\scriptsize{(61--90)} & 22.6\,\scriptsize{(10--39)} & 22.6\,\scriptsize{(10--39)} & 66.9 \\
        LSTM & 12.9\,\scriptsize{(0--26)} & 87.1\,\scriptsize{(74--97)} & 83.9\,\scriptsize{(71--94)} & 6.3 \\
        A3/A5 & 100.0\,\scriptsize{(100--100)} & 0.0\,\scriptsize{(0--0)} & 3.2\,\scriptsize{(0--10)} & 72.4 \\
        \bottomrule
    \end{tabular}
\end{table}


It is important to mention that the ping-pong column is very important operationally in current \ac{ran} deployments and even more important in future (much denser with many overlapping small cells) cloud deployments in which the best cell in terms of signal-strength can vary faster than our test-data. In this case, A3/A5 type rules may have an even higher ping-pong rate. In this regard, we remind that the Zero-Knowledge baseline shares the same \ac{hgt} backbone, recurrent module, and candidate-set scoring function as \ac{ilcp}, but it re-initializes the per-\ac{ue} recurrent state at every \ac{ho}. Its nonzero ping-pong rate therefore isolates the effect of losing cross-\ac{ho} state. The Transformer baseline also produces a higher ping-pong rate. In both cases, the target-side state has not yet rebuilt enough temporal context when a small \ac{rsrp} fluctuation appears after \ac{ho}. In contrast, \ac{ilcp} carries source-side history into the target side and can therefore identify such events as transient rather than immediately \ac{ho}-worthy.
\subsection{Post-handover Cold-Start Gap}
\label{sec:exp-postho}


\begin{figure}[t]
	\centering
	\includegraphics[width=0.95\linewidth]{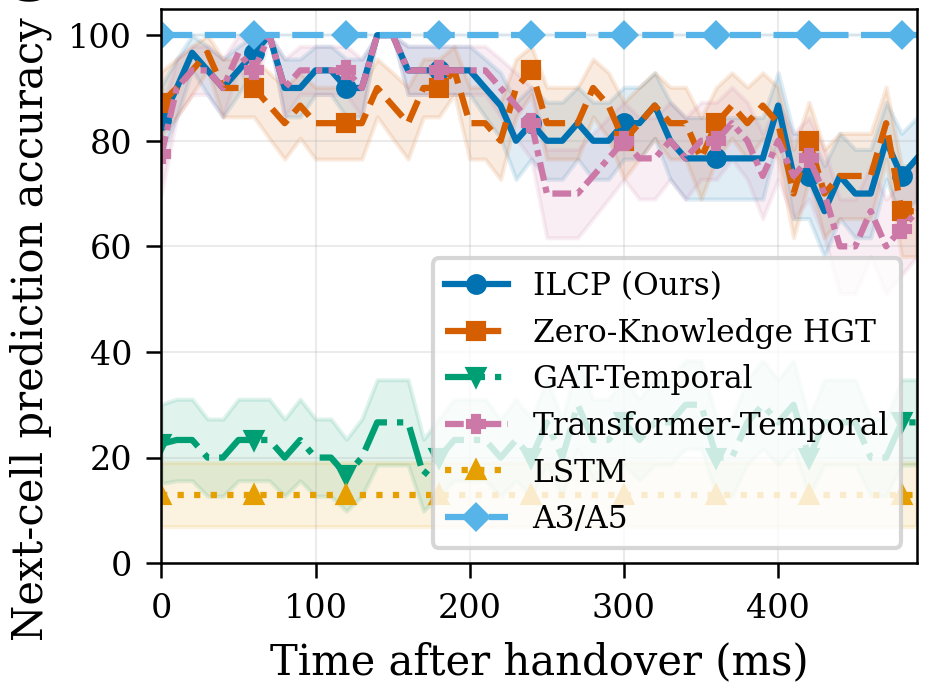}
	\caption{Post-handover next-cell accuracy on Vienna as a function of timesteps after the handover instant, with $1$ step $\approx 10$\,ms.}
	\label{fig:postho}
\end{figure}

Figure~\ref{fig:postho} plots Acc@$\delta$ for $\delta\in[0,30]$ post-handover steps. Each point evaluates the model $\delta$ steps after the handover instant and compares its output with the ground-truth future serving cell at horizon $\Delta$. The Zero-Knowledge baseline and \ac{ilcp} start in the same range at the \ac{ho} instant, but \ac{ilcp} improves over the next few post-handover steps. This is the interval in which the Zero-Knowledge baseline is still rebuilding temporal context from a re-initialized state, while \ac{ilcp} already uses an adapted version of the source-side recurrent state. The gap peaks at $+13.3$\,pp on step 3 and averages $+5.1$\,pp over steps 5 to 25. This measured gap is the empirical manifestation of the cold-start penalty defined in \cref{eq:cold-start-cost}. In operational terms, the result means that the target-side model becomes useful faster after \ac{ho} when source-side recurrent context is preserved. This effect is once again vital for denser deployments with more overlapping cells, where the early post-handover period can contain borderline candidate-cell decisions and short-lived measurement fluctuations. 
Because of the strict page restriction, detailed results about robustness under measurement impairments could not be included within the main body of the paper; instead, they can be found in Appendix~\ref{sec:exp-perturb}.


	\section{Conclusion}
\label{sec:discussion}
%

On the Vienna 4G/5G trace, \ac{ilcp} reduces the cold-start gap relative to the no-transfer baseline and eliminates ping-pongs in the test split. This is very critical for dense deployments with overlapping cells and borderline handover conditions. 
The result suggests that preserving recent source-side measurement and mobility history helps the target side distinguish stable mobility trends from short-lived measurement fluctuations. The comparison with A3/A5 should be interpreted carefully. 
On the unperturbed trace, A3/A5 is strongly aligned with the reference handover labels, so its clean-trace \ac{hof} is not directly comparable to the learned models. The perturbation experiments in Appendix~\ref{sec:exp-perturb} test a different question, i.e., whether a method can recover the fixed reference optimal serving-cell sequence when the input measurements are degraded. Under this setting, robustly trained \ac{ilcp} is less sensitive to shadow fading, blockage, and sparse sampling than the reactive A3/A5 rule. 

The method is lightweight enough for practical deployment. The 128-byte latent payload adds little \ac{xn} overhead. The successful compression indicates that the recurrent state contains a relatively low-dimensional task-relevant subspace. 

There are some limitations which still remain including limitations due to usage of a real-life publicly available network trace. First, the dataset contain a limited number of handover events, so larger traces are needed to validate the transfer mechanism more strongly. Secondly, the perturbation experiments model measurement impairments but do not cover cases where the operationally preferred serving cell truly changes, for example due to an outage or a load shift. Our future work will include controlled simulations comprising varying network topologies, large number of traces, optimal serving cells due to varying criteria, and varying \ac{ue} speeds and mobility patterns. 

	
	\bibliography{main_paper}
	\bibliographystyle{icml2026}
	
	\newpage
	\appendix
	\onecolumn
	\section{List of abbreviations}
\label{sec:appx-abbreviations}

\printacronyms[name=]

	\section{Details on System Design and Experimental Setup}

\subsection{System Diagram}
\label{sec:appx-system-diagram}

\begin{figure*}[h]
    \centering
    \includegraphics[width=0.95\textwidth]{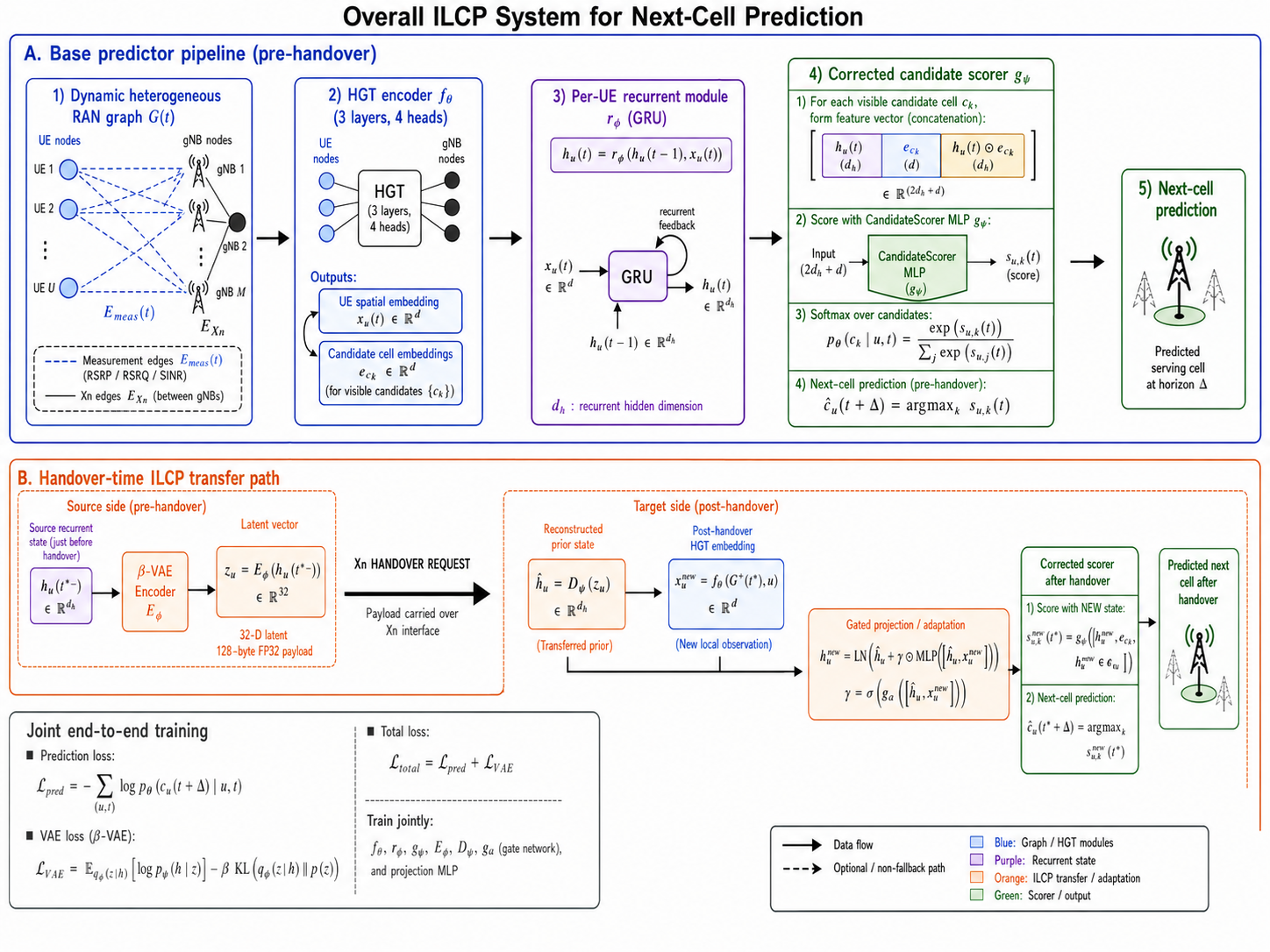}
    \caption{System Diagram}
    \label{fig:system_diagram}
\end{figure*}

Figure~\ref{fig:system_diagram} summarizes the end-to-end ILCP path from the source \emph{g}NB to the target \emph{g}NB at a handover instant: the heterogeneous-graph backbone and GRU produce the pre-handover recurrent state, which the $\beta$-VAE maps to a fixed-size latent carried as an optional information element on the Xn \textsc{Handover Request}.
On the target side, the latent is decoded and fused with the fresh target-side spatial embedding via the gated MLP and LayerNorm block before candidate-set scoring.
Arrows indicate control and data flow through encoding, transport, adaptation, and prediction; dotted boxes group components that share parameters across handovers while remaining logically distinct on source versus target.

\subsection{ILCP inference pseudocode}
\label{sec:appx-pseudocode}

The pseudocode reflects the implementation of our proposed solution. 

\begin{algorithm}[H]
    \caption{ILCP inference at a handover instant $t^\star$}
    \label{alg:ilcp-inference}
    \begin{algorithmic}[1]
        \STATE \textbf{Input:} source-side recurrent state $h_u^{\text{src}}\in\mathbb{R}^{d_h}$ for UE $u$;
            current target-side spatial embedding $x_u^{\text{new}}\in\mathbb{R}^{d_x}$;
            $\beta$-VAE encoder $\mathcal{E}_\phi$, decoder $\mathcal{D}_\psi$;
            gating MLP $g_\theta$; LayerNorm $\mathrm{LN}$;
            visible candidate cells $\{c_1,\dots,c_K\}$ with embeddings
            $\{e_{c_k}\}_{k=1}^K$; candidate-set scorer $g_\psi$.
        \STATE \textbf{Output:} predicted next cell $\hat c_u(t^\star+\Delta)$ and updated state $h_u^{\text{new}}$.
        \STATE
        \STATE \emph{// Source side, at handover decision}
        \STATE $z_u \gets \mathcal{E}_\phi(h_u^{\text{src}})$
            \hfill \COMMENT{32-dimensional latent, $\le 128$ B FP32}
        \STATE Append $z_u$ as an optional IE to the {\sc HANDOVER\,REQUEST}
            on the \ac{xn} interface.
        \STATE
        \STATE \emph{// Target side, on receipt of {\sc HANDOVER\,REQUEST}}
        \STATE $\tilde h_u \gets \mathcal{D}_\psi(z_u)$
        \STATE $\gamma \gets \sigma\!\left(g_\theta\!\left([\tilde h_u, x_u^{\text{new}}]\right)\right)$
            \hfill \COMMENT{learned sigmoid gate}
        \STATE $h_u^{\text{new}} \gets \mathrm{LN}\!\left(\tilde h_u + \gamma \odot \mathrm{MLP}\!\left([\tilde h_u, x_u^{\text{new}}]\right)\right)$
        \STATE
        \STATE \emph{// Candidate-set scoring}
        \FOR{$k=1$ to $K$}
            \STATE $s_k \gets g_\psi^{(\Delta)}\!\left(
                \left[
                    h_u^{\text{new}},\,
                    e_{c_k},\,
                    h_u^{\text{new}}\odot e_{c_k}
                \right]
            \right)$
        \ENDFOR
        \STATE $\hat c_u(t^\star+\Delta) \gets c_{\arg\max_k s_k}$
        \STATE \textbf{return} $\hat c_u(t^\star+\Delta),\, h_u^{\text{new}}$
    \end{algorithmic}
\end{algorithm}

\subsection{Experimental setup and hyperparameters}
\label{sec:appx-setup}

In this section we describe our experimental setup and parameters. 

\textbf{Data split}: The Vienna trajectory traces are partitioned into training, validation, and test trajectories. The split contains 2200 measurement steps over 31 handover events in the test partition, and 25~500 steps over 95 events in the train+val partitions. 

\textbf{Preprocessing}: Per-edge measurement features ($\acs{rsrp}$, $\acs{rsrq}$, $\acs{sinr}$) are normalized per-cell to zero mean and unit variance over the train split, with the normalisation parameters fixed on validation and test. Per-step UE identifiers are anonymized.

\textbf{Hyperparameters}: The hyperparameters are shown in Table~\ref{tab:hparams}.

\begin{table}[H]
    \centering
    \caption{Hyperparameters used for the experiments. The same configuration is used for ILCP and the baselines except where the architecture differs structurally (e.g.,\ Transformer-Temporal replaces the GRU with a single-layer self-attention encoder over a 16-step sliding window).}
    \label{tab:hparams}
    \small
    \setlength{\tabcolsep}{6pt}
    \begin{tabular}{ll}
        \toprule
        Parameter & Value \\
        \midrule
        \multicolumn{2}{l}{\emph{Architecture}} \\
        HGT layers                              & 3 \\
        HGT hidden dim.                         & 128 \\
        HGT attention heads                     & 4 \\
        GRU hidden dim.                         & 128 \\
        $\beta$-VAE latent dim.\ ($d_z$)        & 32 (= 128 B fp32 payload) \\
        $\beta$ (KL weight)                     & 0.001 \\
        Candidate-set size $K$                  & 8 \\
        ILCP gating MLP                          & 2 layers, 128 hidden, ReLU \\
        Inbound dropout (training)              & 0.2 \\
        \midrule
        \multicolumn{2}{l}{\emph{Training}} \\
        Optimizer                               & AdamW \\
        Learning rate                           & $3{\times}10^{-4}$ \\
        Weight decay                            & $10^{-4}$ \\
        Batch size (UEs $\times$ steps)         & $32 \times 16$ \\
        Epochs (max)                            & 80 \\
        Early-stopping patience                 & 8 \\
        Validation split                        & 15\% of trajectories \\
        \midrule
        \multicolumn{2}{l}{\emph{Mixed-perturbation training}} \\
        Mixture weight (clean : impaired)       & 0.5 : 0.5 \\
        Shadow fading $\sigma_s$                & $\sim\!\mathcal{U}\{6, 12\}$\,dB \\
        Quantization step                       & 1 dB \\
        Reporting delay                         & 40 ms \\
        L3 IIR filter coefficient $k$           & 4 (per 3GPP TS 36.331) \\
        \midrule
        \multicolumn{2}{l}{\emph{Evaluation}} \\
        Bootstrap resamples                     & 1000 \\
        CI level                                & 95\% \\
        Inference hardware                      & 1$\times$ NVIDIA GTX~1080 (8\,GB), Intel i7-8700K, 16\,GB RAM \\
        \bottomrule
    \end{tabular}
\end{table}


	\section{Robustness under Measurement Impairments}
\label{sec:exp-perturb}
%
In a real deployment, the measurements used for \ac{ho} decisions may be inaccurate or variable due to shadowing, blockage, reporting delay, filtering, and incomplete measurement sampling. We model these effects as measurement perturbations and evaluate how well each method recovers a fixed reference target (which may be assumed as ``optimal" in some sense) when the input measurements are degraded. This setting is relevant for operational \ac{ran} deployments because \acp{ue} should be handed over to optimal serving cells, rather than to cells that appear preferable only because of short-lived measurement fluctuations. To this end, we perturb the test-split measurements along three axes that commonly affect radio measurements in real deployments. We add shadow fading with $\sigma_s$ from 0 to 12\,dB, including quantization, reporting delay, and the L3 \ac{lte} \ac{rrm} filter. We also add blockage events from 0 to 32 per trace and \ac{ssb}-burst sub-sampling with periods from 1 to 16 steps. As the supervised reference target label, we keep the target fixed to the logged serving cell observed on the original unperturbed reference trace. The experiment therefore measures robustness to measurement perturbations while holding the reference serving-cell sequence fixed. 

\begin{figure}[H]
	\centering
	\includegraphics[width=0.6\linewidth]{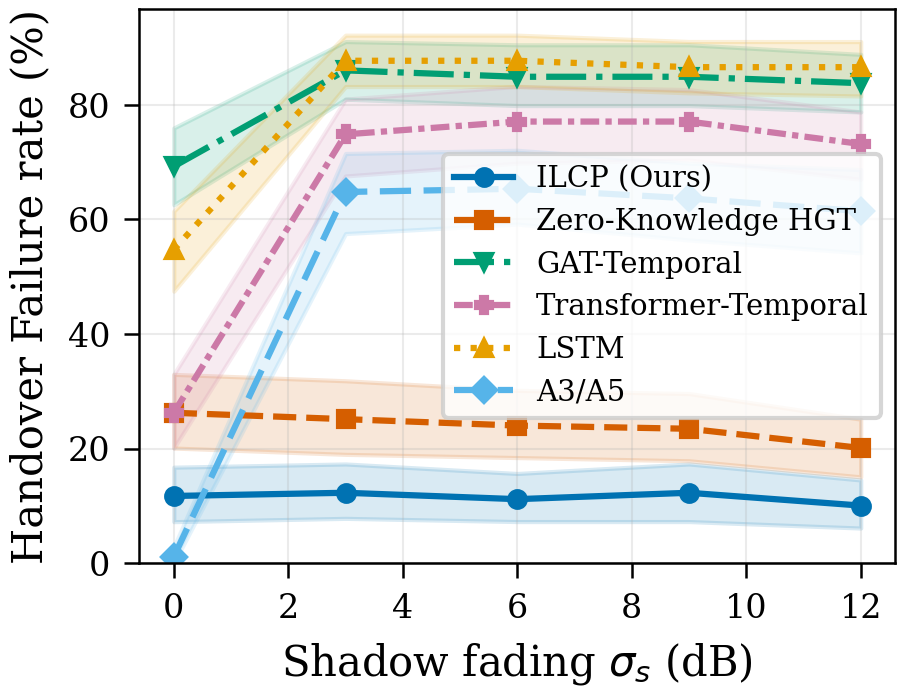}
	\caption{Vienna handover failure rate under shadow-fading perturbations. Robustly trained \ac{ilcp} remains nearly flat across $\sigma_s\in[0,12]$\,dB, while the A3/A5 rule degrades sharply once nonzero shadow fading is added.}
	\label{fig:perturb-noise}
\end{figure}

Table~\ref{tab:perturb-summary} summarizes the change in \ac{hof} from the unperturbed trace to the strongest perturbation setting. Figure~\ref{fig:perturb-noise} shows the shadow-fading sweep on the Vienna 4G/5G data-set. On the unperturbed trace, A3/A5 has very low \ac{hof}, as expected, because the reference handover labels are aligned with the rule-based handover behavior. When shadow fading is added to the measurement stream, its \ac{hof} increases sharply from 1.1\% to the 57--65\% range. Robustly trained \ac{ilcp}, in contrast, remains in the 10--13\% range across the sweep. The Zero-Knowledge baseline is also relatively stable, but at a higher 20--26\% \ac{hof} range.


\begin{table}[t]
    \centering
    \caption{HOF (\%) under realistic measurement impairments: shadow fading $\sigma_s\!\in\!\{0,3,6,9,12\}$\,dB (Noise), \ac{nlos} blockage events $\in\!\{0,4,8,16,32\}$, and \ac{ssb}-burst sub-sampling period $\in\!\{1,2,4,8,16\}$. Each entry: HOF on clean trace $\to$ HOF at the worst level.}
    \label{tab:perturb-summary}
    \small
    \setlength{\tabcolsep}{2.5pt}
    \begin{tabular}{llccc}
        \toprule
        Dataset & Method & Noise & Blockage & SSB \\
        \midrule
        Vienna & \textbf{ILCP (ours)} & 11.7\,$\to$\,12.3 & 11.7\,$\to$\,13.4 & 11.7\,$\to$\,14.5 \\
         & \acs{zk}-HGT & 26.3\,$\to$\,25.1 & 26.3\,$\to$\,27.9 & 26.3\,$\to$\,29.6 \\
         & A3/A5 & 1.1\,$\to$\,65.4 & 1.1\,$\to$\,57.5 & 1.1\,$\to$\,72.6 \\
        \bottomrule
    \end{tabular}
\end{table}

	\section{Bootstrap procedure}
\label{sec:appx-bootstrap}

All confidence intervals reported in the paper are non-parametric percentile-bootstrap \acp{ci} \cite{tibshirani1993introduction} over the test-split handover events. Let $E$ be the set of test-split handover events ($|E|{=}31$ for Vienna). For each of $B{=}1000$ bootstrap rounds we sample $|E|$ events with replacement from $E$, recompute every reported metric on the resampled set, and store the values. The reported point estimate is the mean over the resamples; the lower and upper \ac{ci} edges are the $2.5$th and $97.5$th percentiles. For metrics that are pure proportions (HOF, ping-pong rate, accuracy at a single step) a binomial \ac{ci} would also be appropriate; we use the bootstrap uniformly so that derived metrics (averages over post-handover windows, ratios between methods on the same resample) are estimated under the same procedure.

\end{document}